\documentclass[%
preprint,
superscriptaddress,
nofootinbib,
amsmath,amssymb,
prd,
]{revtex4-1}

\usepackage[english]{babel}
\usepackage[utf8x]{inputenc}
\usepackage[colorinlistoftodos]{todonotes}
\usepackage{enumerate}
\usepackage{graphicx}
\usepackage{dcolumn}
\usepackage{bm}

\usepackage{textcomp}
\usepackage[pagewise,columnwise]{lineno}

\begin{document}

\preprint{APS/123-QED}

\title{
Soft and hard scales of the transverse momentum distribution in the Color String Percolation Model}

\author{J. R. Alvarado Garc\'ia}
\affiliation{Facultad de Ciencias Físico Matemáticas, Benemérita Universidad Autónoma de Puebla, Apartado Postal 165, 72000 Puebla, Puebla, México}

\author{D. Rosales Herrera}
\affiliation{Facultad de Ciencias Físico Matemáticas, Benemérita Universidad Autónoma de Puebla, Apartado Postal 165, 72000 Puebla, Puebla, México}

\author{P. Fierro}
\affiliation{Instituto de Física Luis Rivera Terraza, Benemérita Universidad Autónoma de Puebla, Apartado Postal 165, 72000 Puebla, Puebla, México}

\author{J. E. Ram\'irez}
\email{jhony.ramirezcancino@viep.com.mx}
\affiliation{Centro de Agroecología,
Instituto de Ciencias,
Benemérita Universidad Autónoma de Puebla, Apartado Postal 165, 72000 Puebla, Puebla, M\'exico}

\author{A. Fern\'andez T\'ellez}
\affiliation{Facultad de Ciencias Físico Matemáticas, Benemérita Universidad Autónoma de Puebla, Apartado Postal 165, 72000 Puebla, Puebla, México}

\author{C. Pajares}
\email{pajares@fpaxp1.usc.es}
\affiliation{
Departamento de Física de Partículas and Instituto Galego de Física de Altas Enerxías, Universidad de Santiago de Compostela, E-15782 Santiago de Compostela, España}

\begin{abstract}
In color string models, the transverse momentum distribution (TMD) is obtained through the convolution of the Schwinger mechanism with the string tension fluctuations distribution. Considering a $q$-Gaussian distribution for these fluctuations, the TMD becomes a hypergeometric confluent function that adequately reproduces the characteristic scales at low and high $p_T$ values.
In  this approach, the hard scale of the TMD is a consequence of considering a heavy-tailed distribution for the string tension fluctuations whose width rises as $\sqrt{s}$, multiplicity or centrality increases.
In this paper, we introduce the complete information of the TMD in the color string percolation model by means of the determination of the color suppression factor, which now also depends on the parameters of the $q$-Gaussian.
To this end, we analyze the reported data on pp and AA collisions at different center of mass energies, multiplicities, and centralities.
In particular, for minimum bias pp collisions, we found that the $q$-Gaussian parameters and the effective temperature are monotonically increasing functions of the center of mass energy.
Similar results are found for AA collisions as a function of the centrality at fixed $\sqrt{s}$.
 We summarize these results in a phase diagram that indicates the $q$-Gaussian parameters region allowing the quark-gluon plasma formation.
\end{abstract}
\maketitle


\section{Introduction}

In high energy physics experiments, two projectiles (usually hadrons or nuclei) traveling at relativistic velocities collide.
These projectiles look like thin disks because of the Lorentz contraction\cite{Chaudhuri:2012yt,Busza:2018rrf}. 
Instants before their interaction, color flux tubes are projected into the transverse plane from the partons of each hadron or nuclei in the collision \cite{LAPPI2006200}, which are depicted as small disks carrying color fields, namely color strings, and the fundamental color interaction is given by the string overlapping \cite{Pajares96, Ferreiro2012}.

The number of the color strings in the systems grows, in general, with increasing the center of mass energy, but also with multiplicity for pp collisions,  and size of the colliding objects and centrality for AA collisions \cite{Pajares2005}.
In consequence, strings start to overlap forming clusters in a similar way as in the two-dimensional continuum percolation theory occurs \cite{NARDI199814}.
In this context, the color string percolation model (CSPM) is a suitable framework for analyzing the properties of the colliding systems. Here, the multiparticle production is described in terms of the color strings, which  break and create new strings through neutral color objects, producing latter the observed hadrons.
When the string clusters are formed, they behave as a single color source with a color field corresponding to the vector sum of the individual ones. The main result due to the random oriented color summation of overlapped strings is a reduction of the multiplicity and an increase of the string tension, hence an increase of the mean transverse momentum \cite{Braun:2015eoa, string}.

It is worth mentioning that the CSPM can explain most of the experimental data on pp, pA, and AA collisions, including but not limited to the azimuthal distribution of the produced particles, as well as the temperature dependence of the ratio between the shear and bulk viscosities over the entropy density \cite{bautista2012,BRAUN201314, braunridge,Deus2016,Carlota, Sahoo2019}.
Moreover, it is possible to introduce a parameter like a temperature in the CSPM (and other color string models)  by assuming the Schwinger mechanism for particle production. 
If the string tension fluctuates, the transverse momentum distribution (TMD) can be estimated by the convolution of the Schwinger mechanism with the fluctuations distribution of the string tension.
In particular, if these fluctuations are modeled by a normal distribution, the Schwinger mechanism becomes an exponential decay, the well-known thermal distribution, which constant decay is related to the inverse of the effective temperature \cite{BIALAS1999301, DiasdeDeus:2006xk}. 
Even though, this result has been extensively used, it only adequately describes the TMD at low transverse momentum values.

Usually, the transverse momentum spectrum is fully described by a fit function composed by an exponential decay term plus a decreasing power-like contribution representing the soft non-perturbative and hard perturbative QCD collisions, respectively \cite{BYLINKIN201465, Bylinkin}.
A shortcoming with this approach is the impossibility to identify the string tension fluctuations that originate this TMD after their convolution with the Schwinger mechanism.

The importance of the relation between the hard and soft parts of the TMD have been emphasized recently in connection with the possible fast thermalization of the quark gluon plasma produced in pp and AA collisions. In the same way that in conformal field theory the energy cut-off for the ultraviolet modes sets the effective thermal behavior of the system \cite{Calabrese:2016xau,Berges:2017zws}, a hard parton interaction in a high energy collision produces a rapid quench of the entangled partonic initial state and thus the corresponding effective temperature, inferred from the exponential shape of the TMD. In addition, it was shown that the fluctuations of the hard scale leads to the effective temperature \cite{Baker:2017wtt,Feal:2018ptp,Feal2021}.

On the other hand, in a recent paper \cite{JETMD}, C. Pajares and J. E. Ramírez introduced a TMD obtained by considering that the string tension fluctuations are a $q$-Gaussian distribution instead of a normal distribution. In this way, the full TMD is described by a confluent hypergeometric function, which adequately reproduces the exponential decay at low $p_T$ and also has a power-like behavior at high $p_T$, describing the soft and hard scales of the TMD that only depend on the $q$-Gaussian parameters.
Using this approach it has been possible to fit the TMD of charged hadrons produced in pp collisions at different center of mass energies.
Moreover, the TMD of Higgs bosons reconstructed from $H\to\gamma\gamma$ and $H\to 4l$ decays in pp collisions at $\sqrt{s}$=13 TeV are also well described.
Notably, this confluent hypergeometric function has been derived to describe the TMD by considering complex arguments for string tension fluctuations \cite{Feal2021}.

In this paper, we aim to introduce the soft and hard scales of the transverse momentum distribution into the CSPM. To do this, we explore modifications on the average of the transverse momentum squared and the color suppression factor produced by the introduction of the $q$-Gaussian as the distribution that models the color string fluctuations.

The plan of this paper is as follows. In Sec.~\ref{sec:CSPM}, we present the CSPM and its fundamental phenomenology. In Sec.~\ref{sec:TMD}, we discuss the procedure to obtain the TMD from considering that the string tension fluctuations are modeled by a $q$-Gaussian distribution and how the hard part of the TMD contributes on $\langle p_T^2 \rangle$ and the color suppression factor.
In Sec. \ref{sec:results} we describe the analysis of the experimental data and we show our main results. Finally, Sec.~\ref{sec:conclusions} contains our conclusions and perspectives.

\section{Color String Percolation Model}\label{sec:CSPM}

\begin{figure}[ht!]
    \centering
    \includegraphics[scale=0.1]{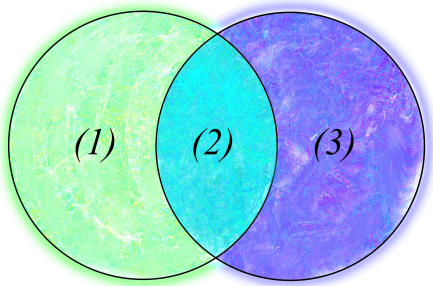}
    \caption{Sketch of the interaction between two color sources.}
    \label{Est}
\end{figure}

The nature of the color fields and their interactions in particle collisions are conveniently described by percolation theory of fully penetrable disks, as we mentioned before.
The color strings are represented in the transverse plane by disks of radius $r \approx$ 0.2-0.3 fm 
\cite{Amelin:1993cs,Braun:1999hv,Braun:2015eoa, string}. 
For simplicity, all the color strings are considered of the same area $S_1$. Each individual color source has a color charge arbitrarily oriented  $\mathbf{Q}_1$ with color field intensity $Q_1$ and color density $\rho_Q = Q_1/S_1$, multiplicity $\mu_1$ and average squared transverse momentum $\langle p_T ^2 \rangle_1$ \cite{Braun:1999hv,Braun:2015eoa, string}.
In this picture, parton interactions are modeled by the overlapping of objects that carry color charge.
Let us discuss an example of a cluster formed of two overlapping strings as we depict in Fig.~\ref{Est}. In this scenario, we have three different color sources: green (1), cyan (2), and blue (3).
Notice that ${(1)}$ and ${(3)}$ are two independent color sources with equal area $S^{(3)} =S^{(1)}$, such that $ S^{(1)} = S_1 - S^{(2)}$, while $S^{(2)}$ is the area of the region (2), where the contributions of color fields sum one another. We assume that each area individually produces charged particles. This means that each area has its own color density, multiplicity, and transverse momentum.
Since the overlapping is partial, the color charge on each region with no overlapping, color sources (1) and (3), is 
\begin{equation}
    Q^{(1)} = \rho_Q S^{(1)} = Q_1 (S^{(1)}/S_1)  .
\end{equation}
Notice that each color source contributes to the color charge on the overlapping region (2) as
\begin{equation}
    \rho_Q S^{(2)} = Q_1 (S^{(2)}/S_1).
\end{equation}
Thus the color charge $Q^{(2)}$ over $S^{(2)}$ is the vector sum considering that color sources (1) and (3) are of equal color field intensity and in average $\langle \mathbf{Q}_1, \mathbf{Q'}_1 \rangle = 0$ \cite{Biro:1984cf,Ferreiro:2003dw,ramirez2021interacting}. 
Therefore
\begin{equation}
    Q^{(2)} = \sqrt{2}\rho_QS^{(2)} = \sqrt{2}Q_1 (S^{(2)}/S_1).
\end{equation}
In the same way, since $\mu^{(j)}$ is proportional to the color charge $Q^{(j)}$ by $\mu^{(j)} = \mu_1 Q^{(j)}/Q_1$ with $j=1,2$,  the total multiplicity can be expressed as  \cite{Braun:1999hv,Braun:2015eoa, string}
\begin{eqnarray}
   \nonumber \mu &=& 2\mu^{(1)} + \mu^{(2)}  \\
    &=& 2 \mu_1 \frac{S^{(1)}}{S_1} + \sqrt{2} \mu_1 \frac{S^{(2)}}{S_1}.
\label{eq:mult2S}
\end{eqnarray}
Similarly, the average squared transverse momentum is  \cite{Braun:1999hv,Braun:2015eoa, string}
\begin{eqnarray}
\nonumber \langle p_T^2 \rangle &=& 2 \langle p_T^2 \rangle^{(1)} + \langle p_T^2 \rangle^{(2)} \\
&=& 2 \frac{\mu_1}{\mu} \frac{S^{(1)}}{S_1}\langle p_T^2 \rangle _1 + \sqrt{2} \frac{\mu_1}{\mu} \sqrt{2} \frac{S^{(2)}}{S_1}\langle p_T^2 \rangle_1.
\label{eq:pTsqr2S}\end{eqnarray}
By using Eq.~\eqref{eq:mult2S} and Eq.~\eqref{eq:pTsqr2S} and taking into account that $S^{(1)}+S^{(2)} =  S_1$ we compute  \cite{Braun:1999hv,Braun:2015eoa, string}
\begin{eqnarray}
    \nonumber \frac{\langle p_T^2 \rangle}{\langle p_T^2 \rangle_1} &=& \frac{2(S^{(1)}/S_1)+2(S^{(2)}/S_1)}{2(S^{(1)}/S_1)+\sqrt{2}(S^{(2)}/S_1)}\\
    &=& \frac{2}{2(S^{(1)}/S_1)+\sqrt{2}(S^{(2)}/S_1)} .
\end{eqnarray}
Generalizing for a cluster of $N$ strings, 
$\mu$ and 
$\langle p_T^2\rangle$
are given by
\begin{eqnarray}
     \mu&=& \sum^{M}_{i} \mu^{(i)} = \sum^{M}_{i}\frac{\sqrt{n_i}{S}^{(i)}}{S_1} \mu_1, 
     \label{eq:mp} \\
         \langle p_T^2\rangle &=&
         \sum^{M}_{i} \langle p_T^2\rangle^{(i)} =
         \frac{{\sum_i}^M\left(\frac{n_i{S}^{(i)}}{S_1}\right)} {{\sum_i}^M \left(\frac{\sqrt{n_i}{S}^{(i)}}{S_1}\right)}  \langle p^2_T\rangle_1 , 
       \label{eq:pm}  
\end{eqnarray}
with $M$ being the total number of generated color sources in the cluster \cite{Braun:1999hv,Braun:2015eoa}. It counts the number of the regions $S^{(i)}$ with $n_i$ overlapped strings. Note that in the case $n_i=1$ corresponds to the remnant fraction surfaces with no overlapping (see Fig.~\ref{Est}).
We use relations \eqref{eq:pm} and $\sum_{i}^{M}n_i S^{(i)} = N S_1$ to establish 
\begin{equation}
    N = \frac{\mu}{\mu_1} \frac{\langle p_T^2 \rangle}{\langle p_T^2 \rangle_1},
\label{eq:conservLaw}
\end{equation}
which suggests a conservation law for the transverse momentum of the produced color sources \cite{Braun:1999hv,Braun:2015eoa, string}.


Notice that Eq.~\eqref{eq:mp} can be split as a sum expressed in terms of the total area $S^{\text{tot}}_n$ defined as the sum of all regions that has $n=n_i$ overlapped disks. Thus,  
\begin{eqnarray}
 \nonumber     \frac{\mu}{\mu_1}
&=&
\nonumber \frac{\sqrt{1}}{S_1}  \sum_{i,n_i=1}  {S}^{(i)}_{n_i=1}
+  \frac{\sqrt{2}}{S_1}  \sum_{i,n_i=2}  {S}^{(i)}_{n_i=2}
+
\cdots
+   \frac{\sqrt{N}}{S_1}  \sum_{i,n_i=N}  {S}^{(i)}_{n_i=N}
\\&=&
\sum_{n=1}^{N}
 \frac{\sqrt{n}}{S_1}  \sum_{i,n_i=n}  {S}^{(i)}_{n_i=n}  = \sum_{n=1}^{N}
 \frac{\sqrt{n}}{S_1}  S_{n}^{\text{tot}}
,
\label{eq:multN}
\end{eqnarray}
and from Eq.\eqref{eq:conservLaw} \cite{Braun:1999hv,Braun:2015eoa, string}
\begin{equation}
    \frac{\langle p^2_T\rangle}{\langle p^2_T\rangle_1}
    = N \frac{\mu_1}{\mu} =
    \frac{N}{ \sum_{n=1}^{N}
 \frac{\sqrt{n}}{S_1}  S_{n}^{\text{tot}}}  .
\end{equation}

Assuming uniformly distributed strings, Eq.~\eqref{eq:multN} becomes
\begin{equation}
\frac{\mu}{\mu_1} 
= \frac{\langle \sqrt{n}\text{ } \rangle}{S_1} S
= N\frac{\langle \sqrt{n} \text{ } \rangle}{\xi} , 
\label{eq:15}
\end{equation}
where $\xi= NS_1/S$ is the string density or filling factor, which describes the occupation of $N$ strings of area $S_1$ over the total interaction area $S$ \cite{Braun:1999hv,Braun:2015eoa, string}.
On the other hand, from Eq. \eqref{eq:15} we can see a damping on multiplicity given by
\begin{equation}
    F(\xi) = \frac{\mu}{N \mu_1} = \frac{\langle \sqrt{n}\text{ } \rangle}{\xi},
\end{equation}
where the average on the number of strings is taken from a Poisson distribution and represents a reduction in multiplicity on a certain number of strings  \cite{Braun:1999hv,Braun:2015eoa, string}. We call this term the color suppression factor $F(\xi)$, which emerges naturally from cluster formation. 
 Notice that $F \leq 1$, because the number of strings times the multiplicity of a single string is higher or equal to the total multiplicity $\mu$.
In the thermodynamic limit $F(\xi)$ is given by  \cite{Braun:1999hv,Braun:2015eoa, string}
\begin{equation}
    \frac{1}{F^2(\xi)} = \frac{\xi}{ 1 - e^{-\xi} } .
\end{equation}
Finally, from Eqs. \eqref{eq:15} and \eqref{eq:conservLaw} we can express multiplicity and squared average transverse momentum as functions of the color suppression factor  \cite{DiasdeDeus:2006xk}
\begin{eqnarray}
\mu & = &  \mu_1 N F(\xi), \\ \label{eq:mu}
\langle p_T^2 \rangle & = & \langle p_T^2 \rangle_1/ F(\xi).
\label{eq:pt}
\end{eqnarray}
In Sec.~\ref{sec:TMD} we discuss the method to compute the color suppression factor by using the information of the transverse momentum distribution. In a general way, to estimate the filling factor $\xi$, it is necessary to solve the equation $F(\xi)=b$, or equivalently
\begin{equation}
    \frac{1-\exp(-\xi)}{\xi}=b^2,
\end{equation}
which solution is given by
\begin{equation}
 \xi=\frac{1}{b^2} + W \left(- \frac{\exp(-1/b^2)}{b^2}  \right),\label{eq:eta}
\end{equation}
where $W$ is the Lambert function.


\section{Soft and hard scales of the TMD}\label{sec:TMD}

In some color string models, the transverse momentum distribution of the particles produced is described by the Schwinger mechanism \cite{schwinger,wong1994introduction}:
\begin{equation}
    \frac{dN}{dp_T^2}\sim \exp \left( -\frac{\pi p_T^2}{x^2}  \right),
    \label{eq:SM}
\end{equation}
where $x$ is the string tension,  which can fluctuate.
If $P(x)$ is the distribution describing the string tension fluctuations, then the TMD is computed as the convolution of the Schwinger mechanism in Eq.~\eqref{eq:SM} with the string tension fluctuations as follows \cite{BIALAS1999301}
\begin{equation}
    \frac{dN}{dp_T^2}\sim\int_0^\infty \exp \left( -\frac{\pi p_T^2}{x^2}  \right) P(x)dx.\label{eq:convolution}
\end{equation}
In particular, if $P(x)$ is a Gaussian distribution centered at 0 with
variance $\sigma^2=\langle x^2 \rangle$, the Schwinger mechanism \eqref{eq:SM} becomes a thermal distribution \cite{BIALAS1999301, DiasdeDeus:2006xk}
\begin{equation}
    \frac{dN}{dp_T^2}\sim \exp\left(  -\beta p_T \right),
    \label{eq:MSthermal}
\end{equation}
where $\beta=(2\pi/\sigma^2)^{1/2}$ can be interpreted as the inverse temperature of the system since Eq.~\eqref{eq:MSthermal} is similar to the Boltzmann distribution \cite{DiasdeDeus:2006xk}. The average of the transverse momentum squared is
\begin{equation}
    \langle p_T^2 \rangle = \frac{\int_0^\infty p_T^2 \exp(-\beta p_T)dp_T}{\int_0^\infty \exp(-\beta p_T)dp_T}=\frac{\sigma^2}{\pi}. \label{eq:pt2gauss}
\end{equation}
We can define a temperature for the color string percolation model by comparing $\langle p_T^2 \rangle$ computed from the Schwinger mechanism \eqref{eq:pt2gauss} and the one deduced from the overlapping of color strings \eqref{eq:pt}, which implies $\sigma^2=\pi \langle p_T^2 \rangle_1/F(\xi)$ \cite{DiasdeDeus:2006xk}. Thus
\begin{equation}
    T(\xi)=\sqrt{\frac{\langle p_T^2 \rangle_1}{2F(\xi)}}.\label{eq:TCSPM}
\end{equation}

Even though Eq.~\eqref{eq:MSthermal} reproduces the soft part of the transverse momentum distribution, it does not describe the hard part.

\begin{figure}[ht]
\centering
\includegraphics[scale=1]{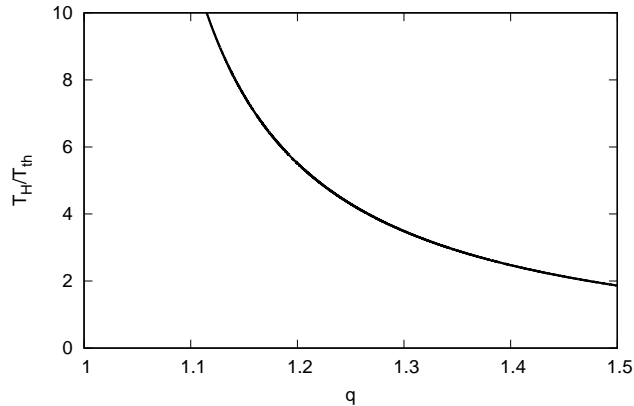}
\caption{Ratio $T_H/T_{th}$ in Eq.~\eqref{eq:ratio} as a function of the parameter $q$.}
\label{fig:Tratio}
\end{figure}

One natural and economical way (by introducing only one new parameter) to describe both scales of the TMD is considering that the string tension fluctuations are described by a $q$-Gaussian distribution \cite{budini} 
\begin{equation}
P(x)=\mathcal{N}\left( 1+\frac{(q-1)x^2}{2\sigma^2}  \right)^\frac{1}{1-q}
\end{equation}
centered at 0 with width of the distribution $\sqrt{\sigma^2}$.
We also assume that $1<q<3$.
Here, $\mathcal{N}$ is the constant that guarantees the normalization of the distribution on the domain of the string tension fluctuation.
Under these considerations and introducing the variable
\begin{equation}
    \tau=\frac{2\sigma^2}{(q-1)x^2},
\end{equation}
the TMD is computed as follows
\begin{equation}
    \frac{dN}{dp_T^2}\sim \frac{\Gamma \left( \frac{1}{q-1} \right)}{\sqrt{\pi} \Gamma\left( \frac{q-3}{2(1-q)} \right)} \int_0^\infty \exp\left( -\frac{\pi p_T^2(q-1)}{2\sigma^2}\tau  \right) \tau^{\frac{1}{q-1}-\frac{3}{2}} (1+\tau)^{\frac{1}{1-q}} d\tau. \label{eq:conv}
\end{equation}
On the other hand, the confluent hypergeometric function is defined as \cite{arfken}
\begin{equation}
    U(a, b, z)=\frac{1}{\Gamma(a)}\int_0^\infty \exp(-zt) t^{a-1}(1+t)^{b-a-1} dt. \label{eq:U}
\end{equation}
Comparing \eqref{eq:conv} and \eqref{eq:U} we identify
\begin{eqnarray}
a =  \frac{1}{q-1}-\frac{1}{2} & \text{ and } & b  =  \frac{1}{2}.
\end{eqnarray}
Thus, we can write the TMD as
\begin{equation}
\frac{dN}{dp_T^2}\sim \frac{1}{\sqrt{\pi}}\Gamma \left( \frac{1}{q-1}  \right) U\left( \frac{1}{q-1}-\frac{1}{2}, \frac{1}{2}, \pi p_T^2 \frac{q-1}{2\sigma^2}  \right), \label{eq:U2}
\end{equation}
where $U$ is the confluent hypergeometric function, which has two well-known asymptotic behaviors \cite{U}.
At low $p_T$,
\begin{equation}
    \frac{dN}{dp_T^2} \sim \exp\left( -\frac{\sqrt{2\pi (q-1)}\Gamma\left( \frac{1}{q-1} \right) p_T}{\Gamma\left( \frac{1}{q-1}-\frac{1}{2} \right) \sigma}  \right). \label{eq:U2thermal}
\end{equation}
Similarly to Eq.~\eqref{eq:MSthermal}, we define the soft scale as
\begin{equation}
    T_{th}=\sigma \frac{\Gamma\left( \frac{1}{q-1}-\frac{1}{2} \right)}{\sqrt{2\pi (q-1)}\Gamma\left( \frac{1}{q-1} \right)}. \label{eq:Tthermal}
\end{equation}
At high $p_T$, the TMD \eqref{eq:U2} behaves as
\begin{equation}
    \frac{dN}{dp_T^2} \sim \frac{\Gamma\left( \frac{1}{q-1} \right)}{\sqrt{\pi}} \left(  \frac{\pi p_T^2 (q-1)}{2\sigma^2} \right)^{\frac{1}{2}-\frac{1}{q-1}}, \label{eq:U2hard}
\end{equation}
and thus we define the hard scale as
\begin{equation}
    T_H= \sigma \sqrt{\frac{2}{\pi(q-1)}} \left( \frac{\sqrt{\pi}}{\Gamma\left( \frac{1}{q-1} \right)}  \right)^{\frac{q-1}{q-3}}. \label{eq:Thard}
\end{equation}
Notice that both the soft and hard scales depend only on the parameters $q$ and $\sigma$ of the $q$-Gaussian distribution.
Moreover, the ratio between the soft and hard scales
\begin{equation}
    \frac{T_H}{T_{th}}=2 \left( \frac{\sqrt{\pi}}{\Gamma\left( \frac{1}{q-1} \right)}  \right)^{\frac{q-1}{q-3}} \frac{\Gamma\left( \frac{1}{q-1} \right)}{\Gamma\left( \frac{1}{q-1}-\frac{1}{2} \right)} \label{eq:ratio}
\end{equation}
only depends on $q$. In Fig.~\ref{fig:Tratio} we depict this ratio.

Notice that $q-1$ marks the departure from the thermal behavior and determines the power like behavior of the TMD at high $p_T$. When $q$ tends to 1 there is not hard part. Later we will show that the normalized fluctuations of the TMD depends only on $q-1$, increasing as $q-1$ does.

To connect this approach to the color string percolation model, we compare the average of the transverse momentum squared computed in both cases. For the TMD obtained by considering the string tension fluctuations as the $q$-Gaussian distribution, we found
\begin{eqnarray}
\langle p_T^2 \rangle &=& \frac{\int_0^\infty p_T^2 U\left( \frac{1}{q-1}-\frac{1}{2}, \frac{1}{2}, \pi p_T^2 \frac{q-1}{2\sigma^2}  \right) dp_T}{\int_0^\infty U\left( \frac{1}{q-1}-\frac{1}{2}, \frac{1}{2}, \pi p_T^2 \frac{q-1}{2\sigma^2}  \right) dp_T} \nonumber\\
 & =&\frac{\sigma^2}{\pi(3-2q)},\label{eq:pt2q}
\end{eqnarray}
for $1<q<3/2$, otherwise $\langle p_T^2 \rangle$ diverges.
The upper bound 3/2 for $q$ has also been reported as the limit for the Bose-Einstein condensation in nonextensive thermodynamics approaches \cite{nonext}.
Comparing Eq.~\eqref{eq:pt2q} with Eq.~\eqref{eq:pt}, the width of the $q$-Gaussian distribution can be related to the color string percolation model through
\begin{equation}
    \sigma^2=\frac{\pi \langle p_T^2 \rangle_1^{qG} (3-2q)}{F(\xi)}.\label{eq:s2}
\end{equation}
Here, we have denoted $\langle p_T^2 \rangle_1^{qG}$ with the superscript $qG$ to emphasize that the mean of the transverse momentum squared of a single string may depend on $q$.
Notice that $\langle p_T^2 \rangle$ rises as $q$ increases (see Eq.~\eqref{eq:pt2q}), which can be done by increasing the string density or enhancing $\langle p_T^2 \rangle_1^{qG}$. 
The former case leads us to an incompatible description of the CSPM with the experimental data.
In what follows, we focus our efforts on the description of the case where increasing the value of $\langle p_T^2 \rangle_1$ correctly reproduces $\langle p_T^2 \rangle$ in Eq.~\eqref{eq:pt2q}.




\begin{figure}[ht]
\centering
\includegraphics[scale=1]{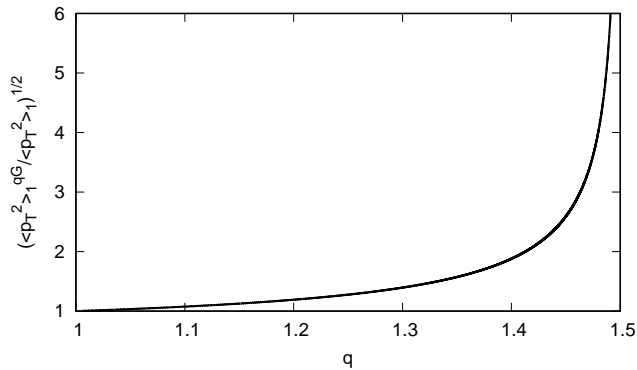}
\caption{Ratio between $\sqrt{\langle p_T^2 \rangle_1^{qG}}$ and $\sqrt{\langle p_T^2 \rangle_1}$ as a function of $q$.}
\label{fig:ptq}
\end{figure}

We expect the analysis of the TMD at low $p_T$ give us a $T_{th}$ value that matches the corresponding $T$ value for the CSPM.
The condition $T_{th}=T$ implies that the color suppression factor is now computed as $F(\xi)=\langle p_T^2 \rangle_1/2T_{th}^2$, and also depends on the $q$-Gaussian parameters.
Moreover, we found that the modified average of the squared transverse momentum of a single string is given by
\begin{equation}
\sqrt{\langle p_T^2 \rangle_1^{qG}}=\sqrt{\langle p_T^2 \rangle_1}\sqrt{\frac{q-1}{3-2q}} \frac{\Gamma \left( \frac{1}{q-1}  \right)}{\Gamma \left( \frac{1}{q-1} -\frac{1}{2} \right)},
\end{equation}
where $\sqrt{\langle p_T^2 \rangle_1}$ is estimated by direct comparison with the value of the critical temperature computed in the lattice QCD framework \cite{DiasdeDeus:2006xk} or the  chemical freeze-out temperature (experimentally determined) \cite{Scharenberg2011}, taking values around 200 MeV.
In Fig.~\ref{fig:ptq} we show the behavior of the quotient $\sqrt{\langle p_T^2 \rangle_1^{qG}}/\sqrt{\langle p_T^2 \rangle_1}$. Notice this case means that the string density coincides with those described by the CSPM. However, the average of the transverse momentum squared of a single string should increase in order to reproduce the hard scale of the TMD. One implication of this approach is the modification of the average of the transverse momentum squared of the CSPM in Eq.~\eqref{eq:pt}, which now reads
\begin{equation}
\langle p_T^2 \rangle= \frac{\langle p_T^2 \rangle_1^{qG}}{F(\xi)}=\frac{\langle p_T^2 \rangle_1}{F(\xi)} \frac{q-1}{3-2q} \left(  \frac{\Gamma \left( \frac{1}{q-1}  \right)}{\Gamma \left( \frac{1}{q-1} -\frac{1}{2} \right)} \right)^2 .
\end{equation}
Clearly this new $\langle p_T^2 \rangle$ is enhanced by the contribution of the hard part of the TMD and it is larger than the predicted value by the CSPM ($\langle p_T^2 \rangle_1/F(\xi)$).



Moreover, the color suppression factor is well determined by the parameters of the $q$-Gaussian distribution, but we still require the value of $\langle p_T^2 \rangle_1$. 

It is interesting to evaluate the variance normalized by the average squared of the string tension fluctuations. To this end, we compute the moments $\langle x^n \rangle=I_n/I_0$, with $I_n$ being the integral
\begin{eqnarray}
I_n &= &\int_0^\infty x^n \left[ 1+\frac{q-1}{2\sigma^2}x^2  \right]^{1/(1-q)} dx \nonumber\\
 & =& \frac{1}{2} \left( \frac{2\sigma^2}{q-1}  \right)^{(n+1)/2} B \left( \frac{1}{q-1} -1-\frac{1}{2}(n-1), 1+\frac{1}{2}(n-1) \right),
\end{eqnarray}
where $B(k_1,k_2)=\Gamma(k_1) \Gamma(k_2)/\Gamma(k_1+k_2)$ is the beta function. The integrals $I_n$ are well defined if $q<(n+3)/(n+1)$.
In particular, the first two moments of the string tension fluctuations are
\begin{eqnarray}
\langle x \rangle &=&\sigma \sqrt{\frac{2}{\pi(q-1)}} \frac{\Gamma \left( \frac{1}{q-1} -1 \right)}{\Gamma \left( \frac{1}{q-1} - \frac{1}{2} \right)},\\
\langle x^2 \rangle & = & \frac{2 \sigma^2}{5-3q}.
\label{eq:VarNorm}
\end{eqnarray}
Thus, the variance normalized by the squared average is given by
\begin{equation}
    \frac{\langle x^2 \rangle-\langle x \rangle^2 }{\langle x \rangle^2}=\frac{\pi (q-1)}{5-3q} \left(  \frac{\Gamma \left( \frac{1}{q-1} -\frac{1}{2} \right)}{\Gamma \left( \frac{1}{q-1} - 1 \right)} \right)^2 -1
\end{equation}
which only depends on $q$. This observable is a bounded monotonic increasing function for $1<q<3/2$, taking values between $\pi/2-1$ and $\pi^2/4-1$.
On the other hand, given the value of the color suppression factor for a particular process, the string density can be computed by using Eq.~\eqref{eq:eta}. 
However, this estimation is adequate for large collisions systems like those produced in heavy ion collisions but may fail to describe pp collisions. In fact, recent studies have pointed out the relevance of finite size effects in all the observables of the CSPM, including the color suppression factor, which significantly deviates from the estimation of the thermodynamic limit \cite{RAMIREZ2017, RAMIREZ2019, Garcia:2022ozz}.



\section{Results}\label{sec:results}

\begin{figure*}[ht!]
\centering
\includegraphics[scale=1]{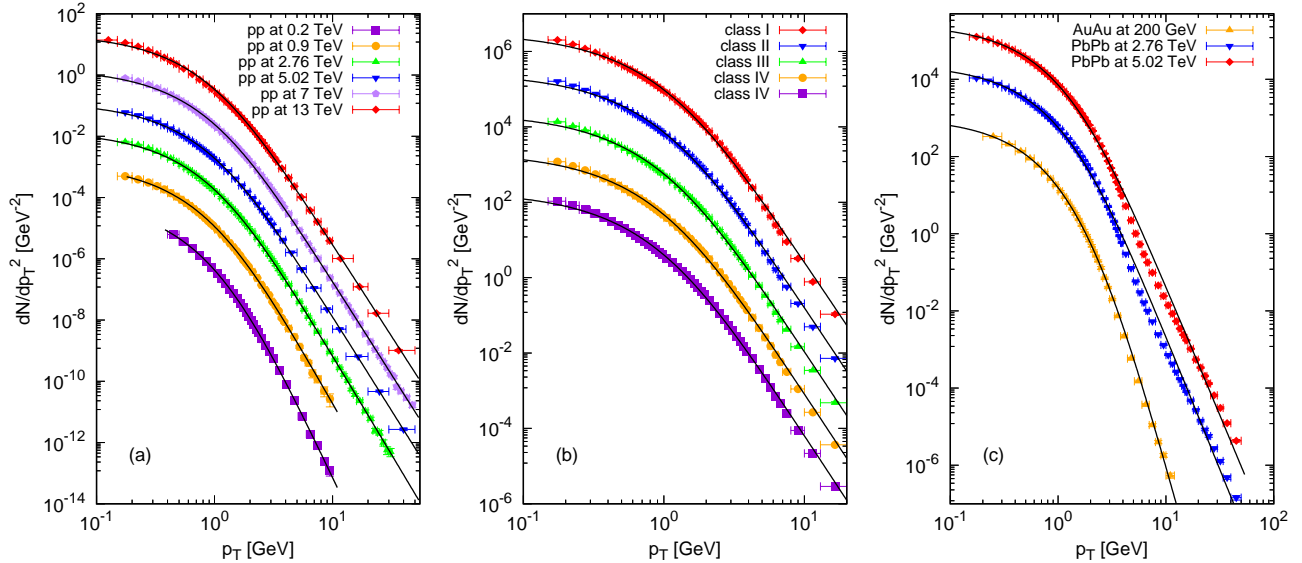}
\caption{Samples of TMD (figures) and their fits (solid lines) for (a) minimum bias pp collisions at different center of mass energies, (b) pp collisions at 13 TeV with multiplicity classification, and (c) central heavy ion collisions at different center of mass energies.}
\label{fig:fits}
\end{figure*}

In this section we discuss our analysis on the experimental data of the production of charged hadrons on pp and AA collisions under different conditions of center of mass energy, centrality, and multiplicity. To this end, we analyse the reported data on Refs. \cite{STAR:2003fka,ALICE:2018vuu,ALICE:2013txf,ALICE:2019dfi,ALICE:2015qqj,ATLAS:2016zkp} with the ROOT 6 software.
In all cases, the confluent hypergeometric function in Eq.~\eqref{eq:U2} is fitted to the TMD data. In this way, we found the values of the $q$-Gaussian parameters $q$ and $\sigma$, and then both the soft and hard scales are computed together with the color suppression factor of the corresponding processes.
In Fig.~\ref{fig:fits} we show a selected sample of TMD and their corresponding fits. 
For the sake of notation, we denote as $dN/dp_T^2$ the invariant yield of charged particles as a function of $p_T$ normalized by the number of events, the pseudorapidity interval, and $2\pi$ corresponding to the azimuthal angle.
To improve visualization, the TMD is scaled by a factor of $10^{L}$ in some instances.
It is worth mentioning that the approach discussed in this manuscript is also valid for fitting the normalized differential primary charged particle cross sections measured at 0.9, 2.76, and 7 TeV, which is proportional to the invariant yield of produced charged particles by the measured luminosity \cite{ALICE:2012fjm} and only differ from the TMD by a multiplicative constant. Moreover, some experiments report a TMD without the $p_T$ normalization. In those cases, the fitting function would be $p_T U$ instead of $U$.

\subsection{Minimum bias pp collisions}

In Fig.~\ref{fig:fits} (a), we show the minimum bias data of pp collisions at different center of mass energies together with their fits according to Eq.~\eqref{eq:U2}.



\begin{figure}[h]
\centering
\includegraphics[scale=1]{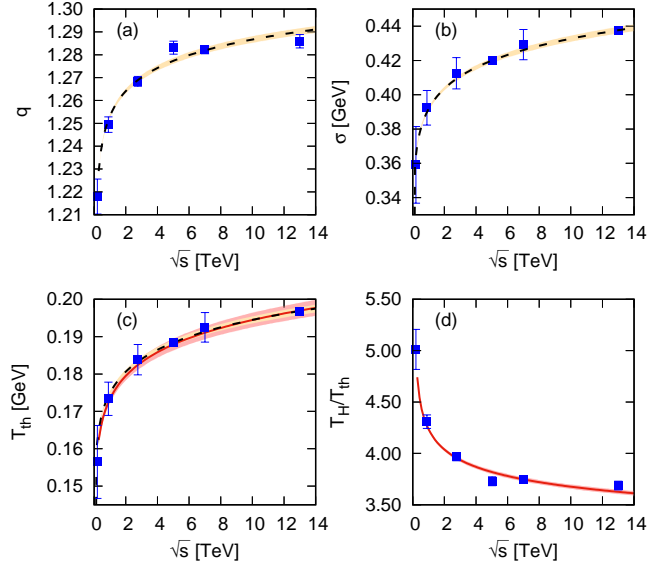}
\caption{(a) $q$ fitted values (squares) as a function of $\sqrt{s}$ and their relation through Eq.~\eqref{eq:qfit} (dashed line).
(b) $\sigma$ fitted values (squares) as a function of $\sqrt{s}$ and their relation through Eq.~\eqref{eq:Sigmafit} (dashed line).
(c) $T_{th}$ estimated (squares) by plugging the $q$ and $\sigma$ fitted values on Eq.~\eqref{eq:Tthermal} as a function of $\sqrt{s}$ and their relation through Eq.~\eqref{eq:Tsoftfit} (dashed line). 
(d) Ratio $T_H/T_{th}$ estimated (squares) by plugging the $q$ and $\sigma$ fitted values on Eq.~\eqref{eq:ratio} as a function of $\sqrt{s}$. 
Red solid lines on (c) and (d) are estimations of $T_{th}$ and $T_{H}/T_{th}$ by using relations $q(\sqrt{s}\text{ })$ and $\sigma(\sqrt{s}\text{ })$ on \eqref{eq:qfit} and \eqref{eq:Sigmafit}, respectively.}
\label{fig:ppminbT}
\end{figure}

For this case, we found that $q$ and the center of mass energy are related as follows
\begin{equation}
q(\sqrt{s}\text{ })=a_q \left( \ln \left( \sqrt{\frac{s}{s_0}}\text{ } \right) -\ln(0.2) \right)^{0.75}+c_q
\label{eq:qfit}
\end{equation}
with $a_q$=0.0247(5), $c_q$=1.218(8), and $\sqrt{s_0}=$1 TeV.
Figure \ref{fig:ppminbT} (a) shows the $q$ fitted data. Notice the agreement between the data and the proposed function $q(\sqrt{s}\text{ })$.

On the other hand, for $\sigma$ and $T_{th}$, we propose as fitting functions
\begin{eqnarray}
{T_{th}}(\sqrt{s}\text{ }) &=& a_{{T_{th}}} \left( \sqrt{\frac{s}{s_0}}\text{ }  \right)^{c_{{T_{th}}}}, \label{eq:Tsoftfit}\\
\sigma(\sqrt{s}\text{ }) &=& a_{\sigma} \left( \sqrt{\frac{s}{s_0}}\text{ }  \right)^{c_{\sigma}}.
\label{eq:Sigmafit}
\end{eqnarray}
We found $a_{T_{th}}=$0.1750(5) GeV and $c_{T_{th}}=$0.046(2), and $a_{\sigma}=$0.392(1) GeV and $c_{\sigma}=$0.043(1).
In Figs.~\ref{fig:ppminbT} (b) and (c) we show the values obtained for $\sigma$ and the value estimated of $T_{th}$, respectively.
In particular, Eq.~\eqref{eq:Tsoftfit} has a power law trend as the previous relation reported in Ref.~\cite{Bylinkin} for $T_{th}$; however, we estimated a slightly lesser value for the exponent.
Moreover, in Fig.~\ref{fig:ppminbT} (c) we also depict the soft scale estimation by plugging into Eq.~\eqref{eq:Tthermal} the functions $q(\sqrt{s}\text{ })$ and $\sigma(\sqrt{s}\text{ })$ from Eqs.~\eqref{eq:qfit} and \eqref{eq:Sigmafit}, finding an excellent match with the function \eqref{eq:Tsoftfit}.

Additionally, $T_H$ is estimated by plugging the fits for $q$ and $T_{th}$ on Eq.~\eqref{eq:ratio} and using the functions $q(\sqrt{s}\text{ })$ and $\sigma(\sqrt{s}\text{ })$. These results are shown in Fig.~\ref{fig:ppminbT} (d).




\subsection{Dependence on the multiplicity of pp collisions}

\begin{figure}[ht]
\centering
\includegraphics[scale=1]{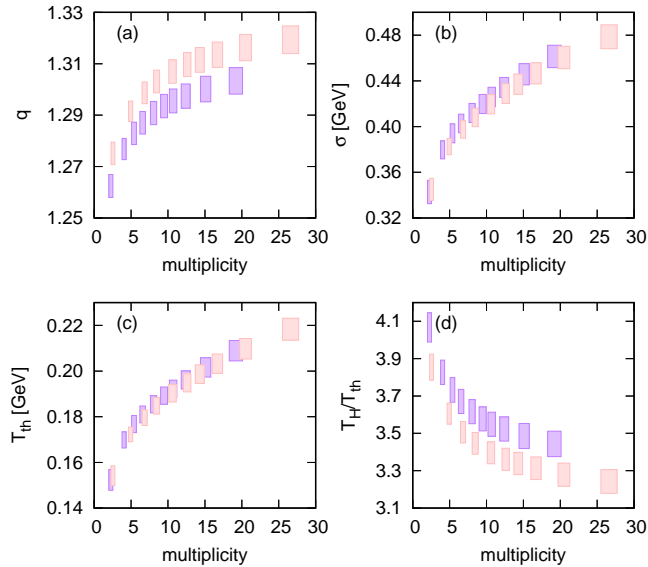}
\caption{Fitted values of (a) $q$ and (b) $\sigma$ for pp collisions at $\sqrt{s}$= 5.02 (purple) and 13 TeV (pink) for the V0M multiplicity classification, and (c) the estimation of the soft scale $T_{th}$, and (d) the ratio $T_H/T_{th}$.}
\label{fig:ppmulT}
\end{figure}

Other TMDs that we can study are those corresponding to the pp collisions with multiplicity classification (see Fig.~\ref{fig:fits} (b)). 
In particular, we analyze the data reported for pp collisions at $\sqrt{s}$=5.02 and 13 TeV by the VZERO ALICE detector, which are classified by the V0M nomenclature.
In Figs.~\ref{fig:ppmulT} (a) and (b) we show the behavior of $q$ and $\sigma$ as a function of multiplicity, respectively. We observe some key features. Both $q$ and $\sigma$ increase as multiplicity and center of mass energy does. Then, the soft scale $T_{th}$ also increases with multiplicity for fixed $\sqrt{s}$ (see Fig.~\ref{fig:ppmulT} (c)).
On the other hand, the hard scale $T_{H}$ can be computed by multiplying $T_{th}$ by the quotient $T_H/T_{th}$ in Eq.~\eqref{eq:ratio}, which only depends on $q$ and it is a decreasing function. 


\subsection{Heavy ions collisions}

\begin{figure}[ht]
\centering
\includegraphics[scale=1]{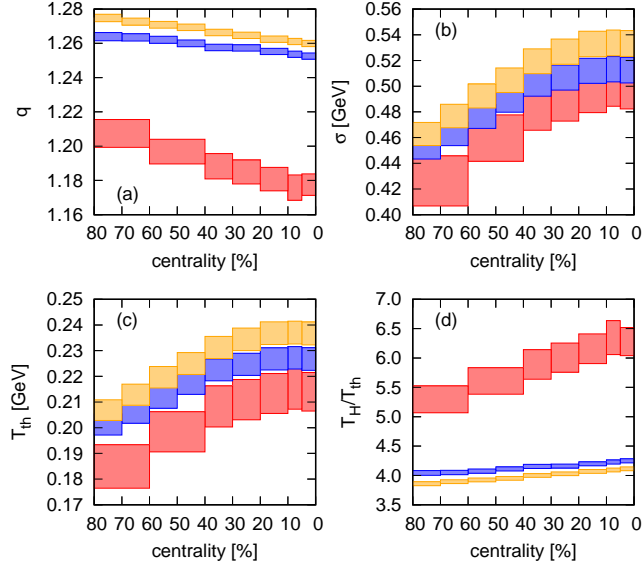}
\caption{Fitted values of (a) $q$ and (b) $\sigma$ for AuAu collisions at $\sqrt{s}$= 0.2 TeV (red), PbPb collisions at 2.76 (blue) and 5.02 (orange) TeV, (c)  estimation of $T_{th}$ and (d) ratio $T_H/T_{th}$, as functions of centrality.}
\label{fig:AAcenT}
\end{figure}

In principle, the TMD reported for heavy ion collision can also be described by using the confluent hypergeometric function in Eq.~\eqref{eq:U2}.
This approach gets to describe the soft and hard parts of the TMD but deviates from the spectra at mesoscale values of $p_T$ (see Fig.~\ref{fig:fits} (c)).
It happens because the final state interactions with a high density medium, the high $p_T$ particles are suppressed and induce the jet quenching \cite{Pajares2005}. Then we fit lower values of $q$ for the most central heavy ion collisions, as it is plotted in Fig.~\ref{fig:AAcenT} (a).
We observe that those deviations are less important as the centrality decreases and the momentum spectra look similar to the TMD of pp collisions.
Nevertheless, the soft scale $T_{th}$ increases as centrality and center of mass energy does, as expected. Our results for $\sigma$ and $T_{th}$ have the same behavior as those for pp collisions. The dependence of $\sigma$ on the energy, the centrality, or the multiplicity is now codified into $F(\xi)$ via Eq.~\eqref{eq:s2}. As energy or centrality (multiplicity) increases,  $F(\xi)$ decreases and $\sigma$ increases. Similar behavior is held for $T_{th}$ because its value is mainly determined by $\sigma$, see Eq.~\eqref{eq:Tthermal}. In pp collisions as the energy or the multiplicity increases, there are more hard collisions and therefore $q$ increases. This is not the case of heavy ion collisions where is well known the suppression of the production of high $p_T$ particles and jet quenching due to the interaction with the quark gluon plasma produced, thus, $q$ decreases. 

Above the percolation threshold of the string density a cluster of the produced strings is formed covering most of the surface of the collision. The diversity of string tensions would be small because there is only one large cluster and a few others with small number of strings. In particular, we observe that the normalized variance decreases with centrality. This is clearly inferred from Fig.~\ref{fig:AAcenT} (a). On the other hand, for the case of pp collisions, the large number of hard collisions produces an increment on $q$, and then an increment of the normalized variance. This different behavior between pp and AA collisions can be traced back to the absence of jet quenching in pp high multiplicity collisions, in contrast to the heavy ion case.

\subsection{Phase diagram for QGP formation}\label{ss:diagram}

\begin{figure}[ht]
\centering
\includegraphics[scale=1]{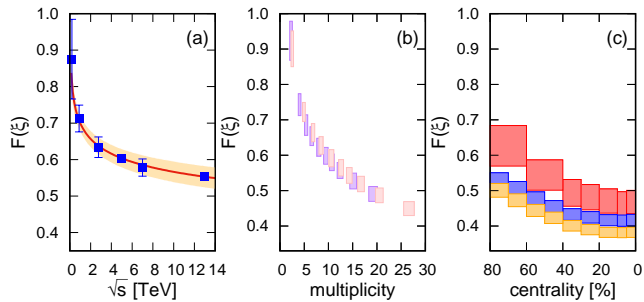}
\caption{Color suppression factor for the cases analyzed: (a) minimum bias pp collisions, (b) pp collisions at $\sqrt{s}$= 5.02 and 13 TeV with multiplicity classification, and (c) heavy ion collisions. Figures, lines, and colors are the same as in Figs.~\ref{fig:ppminbT}, \ref{fig:ppmulT}, and \ref{fig:AAcenT}, respectively.}
\label{fig:Ftodos}
\end{figure}

As we stated in Sec.~\ref{sec:TMD}, the computation of the color suppression factor is now straightforward. For the value of $\sqrt{\langle p_T^2 \rangle_1}$, we consider the estimation using the chemical-freeze out temperature, taking the value 0.207 GeV \cite{Scharenberg2011}. 
Thus, the color suppression factor is calculated as
\begin{equation}
F(\xi)=\frac{0.0214245 \text{ GeV}^2}{T_{th}^2} ,
\label{eq:FTh}
\end{equation}
where $T_{th}$ is the estimated soft scale from the $q$-Gaussian parameters.
In Fig.~\ref{fig:Ftodos}, we show our evaluations of $F(\xi)$. In all cases, the higher values of $F(\xi)$, the lower string density the systems will have. Particularly, the string density rises with the increasing center of mass energy, multiplicity, or centrality, as expected.

\begin{figure}[ht]
\centering
\includegraphics[scale=1]{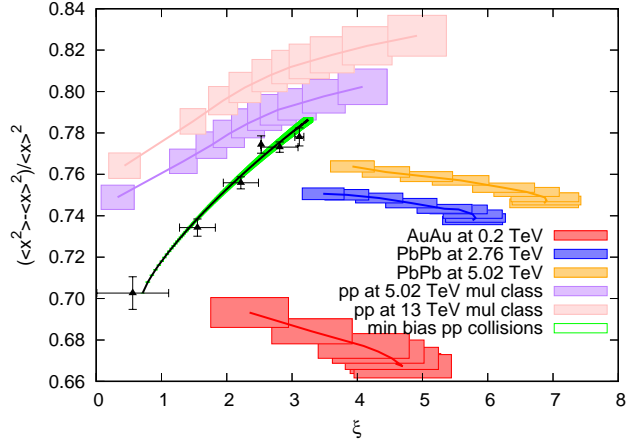}
\caption{Variance normalized by the average square of the string tension fluctuations as a function of the filling factor of the system for the TMD analyzed.}
\label{fig:eta_vs_var}
\end{figure}

In Fig.~\ref{fig:eta_vs_var}, we show the normalized variance as a function of the string density. For all cases, the variance increases as the filling factor grows. 
In particular, the normalized variance for pp collisions has higher values for the higher center of mass energies or multiplicities; meanwhile, it decreases with growing centrality for heavy ion collisions. 

\begin{figure}[ht]
\centering
\includegraphics[scale=1]{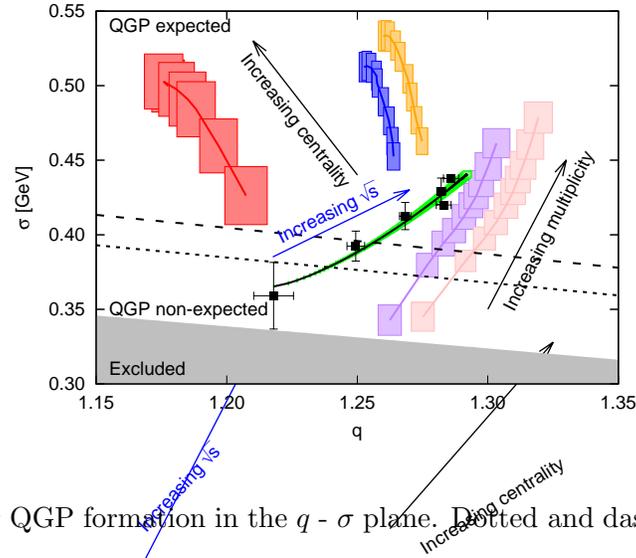}
\caption{Diagram for QGP formation in the $q$ - $\sigma$ plane. Dotted and dashed lines are the critical pairs $q$ - $\sigma$ allowing the QGP formation estimated in the thermodynamic limit and considering finite size effects for pp collisions, respectively.}
\label{fig:Diagram}
\end{figure}

To close this section, let us extend the discussion on the implication of the relation between the color suppression factor and the soft scale on Eq.~\eqref{eq:FTh}. In this way, $F(\xi)$ is also related to the $q$-Gaussian parameters as mentioned before. 
Then, the analyzed LHC and RHIC data (at mid-rapidity region) can be placed in the $q$ - $\sigma$ plane.
We can infer some relevant implications.
When no strings are overlapping, the color field is not suppressed, and $F(\xi)=1$, which occurs for very low dense systems.
Under these conditions, the minimal soft temperature physically acceptable is $T_{th,\text{min}}$=0.146 GeV.
This leads to an exclusion region of forbidden $q$ - $\sigma$ pairs, depicted as a grey shaded region in Fig.~\ref{fig:Diagram}. 
Another noteworthy situation happens when the system reaches the percolation threshold.
At this point, the spanning cluster of color strings emerges, which marks the onset of the QGP in high energy collisions.
The best estimation for the critical value of the color suppression factor in the thermodynamic limit is $F_c=$0.77430816(4) \cite{Garcia:2022ozz}. 
Then, the $q$ - $\sigma$ pairs that give soft scales greater than the critical temperature $T_{th,c}=$0.16634... GeV assures the formation of the QGP for heavy ion collisions.
The critical curve $T_{th,c}=T_{th}(q,\sigma)$ is represented as a dotted line in Fig.~\ref{fig:Diagram}.
On the other hand, for small systems, the finite-size effects on $F_c$ have been found to be $F_c-F_{cL}\propto L^{-1.3}$, giving the value $F_{cL}\approx$0.7066 for pp collisions \cite{Garcia:2022ozz}. The latter leads to a shift in the critical temperature for the QGP formation. Under these conditions, $T_{cL}\approx 0.174...$ GeV for pp collisions. 
Thus, small systems require more energetic collisions to form the QGP than heavy ion collisions.
The critical curve $T_{th,c,L}=T_{th}(q,\sigma)$ for pp collisions is represented as a dashed line in Fig.~\ref{fig:Diagram}.

\section{Conclusions}\label{sec:conclusions}

We presented a natural extension of the color string percolation model to include all the information of the transverse momentum distribution. 
To this end, we have explored the possibility of having string tension fluctuations distributed according to a $q$-Gaussian distribution instead of a normal one.
After the convolution of the Schwinger mechanism with the string tension fluctuations we found that the TMD becomes a confluent hypergeometric function, which can adequately fit the TMD experimental data reported for the production of charged hadrons in pp and heavy ion collisions under a wide range of conditions of center of mass energy, multiplicity classification or (when applicable) centrality.
It is worth mentioning that the confluent hypergeometric function adequately reproduces the expected characteristics of the TMD: exponential decay and power-like behaviors at low and high $p_T$ values, respectively. Then, the soft and hard scales of the TMD have natural definitions.

To connect with the color string percolation model, we compared the prediction of $\langle p_T^2 \rangle$ by using both approaches.
We observe that the incorporation of the hard scale of the TMD rises the value of $\langle p_T^2 \rangle$, 
which can be possible by increasing the value of $\langle p_T^2 \rangle_1$. 
It means that the collision system has the same string density as these modeled by the CSPM, but $\langle p_T^2 \rangle_1$ grows as $q$ takes higher values.
In particular, notice that the ratio $\sqrt{\langle p_T^2 \rangle_1^{qG}}/\sqrt{\langle p_T^2 \rangle_1}$ takes values between 1.19 and 1.4 for the typical values of $q$ and $\sigma$ obtained by fitting the TMD of the processes discussed above.

As we pointed out in Sec.~\ref{ss:diagram}, now the color suppression factor is straightforwardly computed by its relation with the soft scale, given by Eq.~\eqref{eq:FTh}, which also depends on the parameters of the $q$-Gaussian.
In this way, we have introduced the soft and hard scales of the TDM into the color string percolation model.

The introduction of the $q$-Gaussian fluctuations allow us to relate directly the power like hard TMD with its normalized fluctuations, both depend only on $q$. This fact is in line with the possibility that a hard collision produces a rapid quench in the entangled initial partonic state, giving rise to an exponential behavior at low $p_T$ and subsequently to the thermal temperature. 

We condensed all the values obtained for $q$ and $\sigma$ from fits in the $q$ - $\sigma$ plane. By analyzing the characteristic values of the color suppression factor, we found that the low-density limit produces an excluded region of forbidden $q$ - $\sigma$ pairs (shaded region in Fig.~\ref{fig:Diagram}). 
Another notable value is the critical color suppression factor.
The picture of the CSPM in the thermodynamic limit is adequate to describe heavy ion collision. 
In the percolation threshold, the $F_c$ dictates the values of the $q$ - $\sigma$ pairs that mark the departure for QGP formation (see Fig.~\ref{fig:Diagram}). 
Particularly, our results for AuAu collisions at RHIC energies agree with the claim of the QGP observed in those experiments.
On the other hand, for small systems, it is necessary to take into account the finite-size effects, leading to an increment of the critical soft temperature for pp collisions which is consistent with previous estimations.
Notice that minimum bias pp collisions require center of mass energies above 2 TeV to expect the QGP formation.
On the contrary, the QGP formation is not expected for pp collisions with a small production of charged particles, as we showed in Fig.~\ref{fig:Diagram}.

This work can be extended in several ways. 
For instance, it is possible to analyze the TMD of pp collisions in producing a particular charged hadron. This result could give insight into the hardness of the collisions required for producing such charged hadron.
It will be of interest if different $q$ values are observed for different charged hadrons. 
Moreover, since the $q$ and $\sigma$ parameters are involved in determining the color suppression factor, modifications on all observables of the CSPM are expected. 
Finally, we must emphasize that the results presented here are also applicable to study TMDs far from mid-rapidity region, which corresponds to non vanishing baryon-chemical potential.

\begin{acknowledgments}
C. P. thanks the grant Maria de Maeztu Unit of Excellence under the project MDM-2016 0682 of Ministry of Science and Innovation of Spain. This work has been funded by the projects PID2020-119632GB-100 of the Spanish Research Agency, Centro Singular de Galicia 2019-2022 of Xunta de Galicia and the ERDF of the European Union.
This work was funded by Consejo Nacional  de Ciencia y Tecnología (CONACyT-México) under the project CF-2019/2042,
graduated fellowships grant numbers 645654, 1140160, and 848955, and postdoctoral fellowship grant number 289198.
\end{acknowledgments}

\nocite{Calabrese:2016xau,Berges:2017zws,Baker:2017wtt,Feal:2018ptp}

\bibliography{bib}

\end{document}